\documentstyle[epsf,prb,aps]{revtex}
\begin{document}
\draft
\title{HYBRIDIZATION OF ELECTRON SUBBANDS IN A DOUBLE QUANTUM WELL AT
QUANTIZING MAGNETIC FIELD}
\author{V.T.~Dolgopolov, A.A.~Shashkin, E.V.~Deviatov}
\address{Institute of Solid State Physics, Chernogolovka, Moscow
District 142432, Russia}
\author{F.~Hastreiter, M.~Hartung, A.~Wixforth}
\address{Ludwig-Maximilians-Universit\"at, Geschwister-Scholl-Platz
1, D-80539 M\"unchen, Germany}
\author{K.L.~Campman, A.C.~Gossard}
\address{Materials Department and Center for Quantized Electronic
Structures, University of California, Santa Barbara, California
93106, USA}
\maketitle

\begin{abstract}
We employ magnetocapacitance and far-infrared spectroscopy techniques
to study the spectrum of the double-layer electron system in a
parabolic quantum well with a narrow tunnel barrier in the centre.
For gate-bias-controlled asymmetric electron density distributions in
this soft two-subband system we observe both individual subband gaps
and double layer gaps at integer filling factor $\nu$. The bilayer
gaps are shown to be either trivial common for two subbands or caused
by hybridization of electron subbands in magnetic field. We describe
the observed hybrid gaps at $\nu=1$ and $\nu=2$ within a simple model
for the modified bilayer spectrum.
\end{abstract}

\section{Introduction}
\label{intro}

The remarkable properties of a double electron layer in quantizing
magnetic field are determined not only by the relationship between
the intralayer and interlayer Coulomb energies and the
symmetric-antisymmetric level splitting caused by tunneling
\cite{chak,yosh,eis,suen,murphy,lay} but also by the ratio of the
distance between the subbands' electron density weight centres to the
effective Bohr radius \cite{gold}

\begin{equation}
\alpha=d/a_B .\label{eq1}\end{equation}
Throughout the paper we will bear in mind that the meaning of
individual layer is correct to introduce if the tunneling is small;
otherwise, it only makes sense to speak about subbands with
complicated electron density distributions in a double layer system.
So, the interlayer Coulomb correlations were shown to be responsible
for the fractional quantum Hall effect at filling factor $\nu=1/2$
\cite{chak,yosh,eis,suen}, the many-body quantum Hall plateau at
$\nu=1$ \cite{murphy,lay} and the saturation of the phonon-induced
electron drag \cite{gram} with lowering temperature \cite{lilly}.
Besides, these were argued to destroy symmetric-antisymmetric
splitting in strong magnetic fields \cite{boeb,suen1,mac,brey,he}.

It is the ratio $\alpha$ that describes the softness of a bilayer
electron system, i.e. the sensitivity of the subband spacing to
intersubband electron transfer \cite{dol}. From this standpoint, the
balanced double layer with symmetric electron density distributions,
in which the subband electron density weight centres are coincident,
is similar to a conventional two-subband electron system with
vanishing $\alpha$, such as the one in single heterojunctions.
Oppositely, the unbalanced double layer system with asymmetric
electron density distributions is normally soft, i.e. $\alpha\agt 1$,
and so its spectrum is very sensitive to intersubband charge
transfer. In experiment \cite{davies}, e.g., the Landau level fan
chart peculiarities observed at relatively high filling factors in an
unbalanced bilayer system were interpreted in terms of intersubband
charge transfer without appealing exchange and correlation effects.
Moreover, observation has been reported of hybrid gaps at filling
factors $\nu=1,2$ at imbalance \cite{dol}. The very different
behaviours of the gap at $\nu=2$ at imbalance over different ranges
of bilayer electron densities have been detected and associated with
a new phase transition \cite{saw}. A broken-symmetry state in the
fractional quantum Hall regime, whose formation is accompanied by
intersubband charge transfer, has been observed in a wide quantum
well \cite{manoh}. As was discussed in Ref.~\cite{jung}, the simplest
effect of electron correlations is to change the partitioning of
charge between two subbands/layers. It is interesting that the
symmetry breaking does not necessarily imply nonzero charge transfer
between subbands. At filling factor $\nu=2$ at balance, the electron
correlations are expected to break the symmetry of the wave functions
of electron subbands without causing intersubband charge transfer,
which, in particular, points to the possible existence of a "canted
antiferromagnetic" state \cite{zheng}.

Here, using capacitive and far-infrared spectroscopy methods, we
investigate the spectrum of two-dimensional electrons at a quantizing
magnetic field in a parabolic quantum well that contains a narrow
tunnel barrier for the electron systems on either side. Analysis of
the behaviour of the double layer gaps at filling factors $\nu=1$ and
$\nu=2$ at imbalance controlled by gate bias establishes their hybrid
origin as caused by intersubband electron transfer in magnetic field.
A proposed model for magnetic-field-induced hybridization of electron
subbands largely allows description of the experimental results as
well as recent data obtained elsewhere.

\section{Samples and experimental technique}
\label{tech}

The samples are grown by molecular beam epitaxy on semi-insulating
GaAs substrate. The active layers form a 760~\AA\ wide parabolic
well. In the center of the well a 3 monolayer thick
Al$_x$Ga$_{1-x}$As ($x=0.3$) sheet is grown which serves as a tunnel
barrier between both parts on either side. The symmetrically doped
well is capped by 600~\AA\ AlGaAs and 40~\AA\ GaAs layers. The
samples have two ohmic contacts (each of them is connected to both
electron systems in two parts of the well) and two gates on the
crystal surface with areas $120\times 120$ and $220\times 120$
$\mu$m$^2$. The presence of the gate electrode enables us both to
tune the carrier density in the well, which is equal to $4.2\times
10^{11}$ cm$^{-2}$ without gate bias, and to measure the capacitance
between the gate and the well. For capacitance measurements we apply
an ac voltage $V_{ac}=2.4$~mV at frequencies $f$ in the range 3 to
600~Hz between the well and the gate and measure both current
components as a function of gate bias $V_g$, using a home-made $I-V$
converter and a standard lock-in technique. For far-infrared
spectroscopy the sample design is a bit different: a 6~$\mu$m period
Ag grating coupler on the semi-transparent gate is used to couple the
normally incident FIR radiation to the intersubband modes. Our
measurements are performed in the temperature interval between 30~mK
and 2.5~K at magnetic fields of up to 16~T.

The imaginary current component in the low-frequency limit reflects
the thermodynamic density of states in a double layer system at
quantizing magnetic field. This limit occurs if the sample dimension
$L$ is small compared to the current overflowing length
$L_0=(\sigma_{xx}/\pi fC)^{1/2}$, where $\sigma_{xx}$ is the
dissipative conductivity of the bilayer system and $C$ is the
capacitance between gate and quantum well per unit area. The
imaginary current component minima are accompanied by peaks in the
active current component which are proportional at low frequencies to
$(fC)^2\sigma_{xx}^{-1}$ and can be used for measurements of
$\sigma_{xx}$. In the high-frequency limit $L\gg L_0$ both components
of the current tend to zero. Except for the balance point, our
samples are similar to a conventional three electrode system (see,
e.g., Refs.~\cite{ash,dens}), i.e., at imbalance the thermodynamic
density of states is measured in the subband whose electron density
maximum is closer to the gate. If the conductivity of the other
subband does not vanish, the low-frequency limit is reached in the
frequency range used. In this case one can expect weakly
temperature-dependent relatively shallow minima in the sample
capacitance that correspond to gaps in the spectrum of the former
subband. Alternatively, if the Fermi level lies in a gap of the
bilayer spectrum, at lowest temperatures the high-frequency limit is
normally realized. This is indicated by deep minima in the imaginary
current component that depend strongly on temperature. Hence, one can
easily discriminate between one subband gaps and gaps in the bilayer
spectrum.

\section{Results}
\label{results}

\begin{figure}
\vspace{\fill}
\centerline{
\epsfysize=15cm
\epsffile{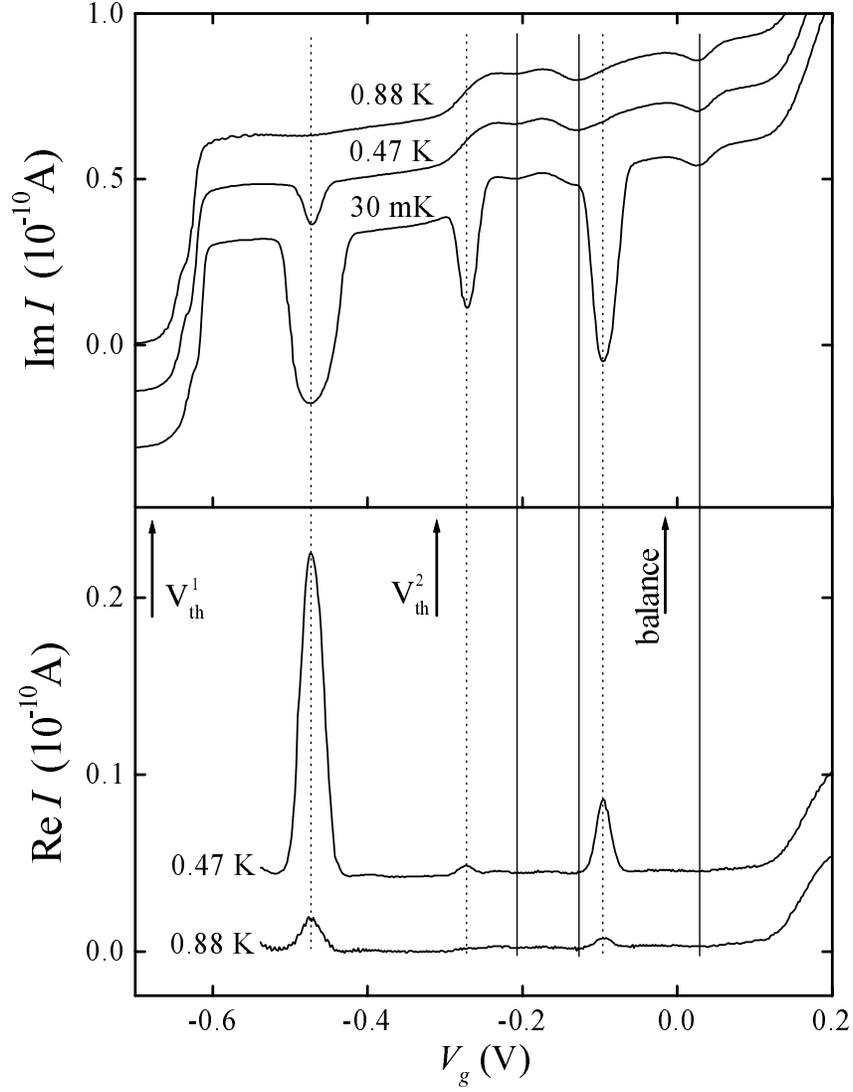}
}
\caption{Dependence of both current components on gate voltage at
$f=300$~Hz at different temperatures in a magnetic field of 2.5~T.
The lines for 30~mK and 0.47~K are shifted for clarity. The vertical
solid (dashed) lines mark minima of the density of states in the
second subband (the system conductivity). The positions of the
subband thresholds and of the balance point are determined from the
fan diagram in Fig.~\protect\ref{fan}.\label{exp}}
\end{figure}

The dependences of both current components on gate voltage in a
magnetic field of 2.5~T at different temperatures are represented in
Fig.~\ref{exp}. At $V_{th}^1< V_g< V_{th}^2$ one subband is filled
with electrons in the back part of the well, with respect to the
gate. At the threshold voltage $V_{th}^2$ a second subband starts to
collect electrons in the front part of the well, as indicated by a
jump of the capacitance. Positions of the system conductivity minima
are marked in the figure by dashed vertical lines. The solid lines
mark minima of the thermodynamic density of states in the second
electron subband. As seen from Fig.~\ref{exp}, the capacitance minima
related to $\sigma_{xx}$ are strongly temperature-dependent, whereas
the measured capacitance in between the deep minima depends weakly on
temperature. In the two-subband case the two kinds of minima co-exist
and, at close positions, switch with varying temperature because of
very different temperature dependences. These determine two Landau
level fans in the ($B,V_g$) plane for the unbalanced double layer
system.

\begin{figure}
\epsfysize=17cm
\epsffile{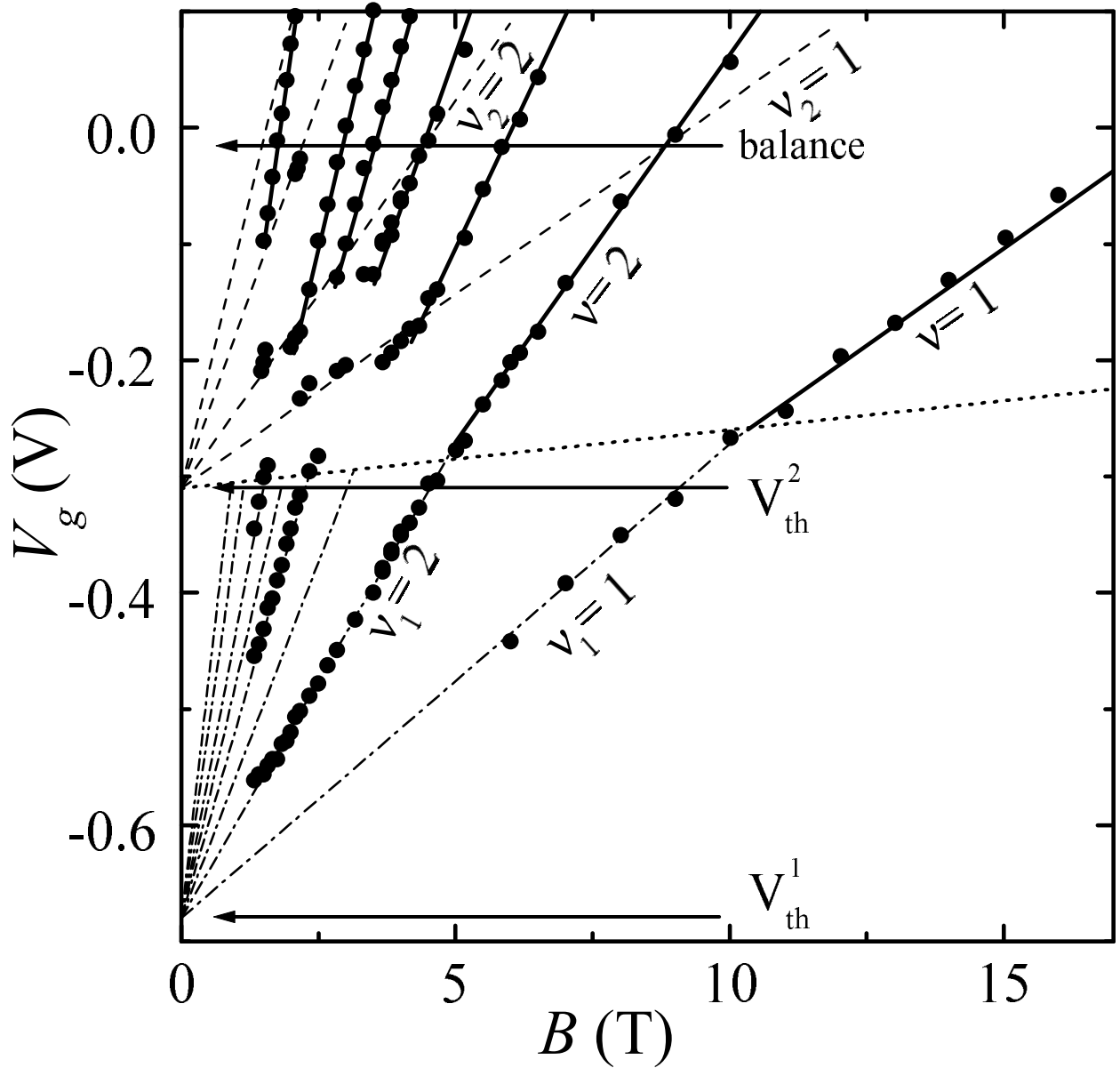}
\caption{Landau level fan chart as found from the minima of the
density of states in the second electron subband at $\nu_2=1,2,4,6$
(dashed lines), of the conductivity in the first subband at
$\nu_1=1,2,4,6$ (dash-dotted lines) and of the double layer
conductivity at $\nu=1,2,3,4,5,6,8,10$ (solid lines). The change of
fan line slopes occurs above the threshold $V_{th}^2$ as marked by
dotted line.\label{fan}}
\end{figure}

Fig.~\ref{fan} presents a Landau level fan chart for our sample. One
subband fans, i.e., that correspond to individual electron subbands,
are shown by dashed and dash-dotted lines, respectively. These are
defined by minima of the thermodynamic density of states in the
second subband at integer filling factor $\nu_2$ (here the
experimental data points are not indicated to avoid overcomplicating
the figure) and of the conductivity in the first subband at integer
$\nu_1$. Their line slopes are inversely proportional to the
capacitance values before and after the jump near $V_g=V_{th}^2$
(Fig.~\ref{exp}). The double layer conductivity minima at integer
filling factor $\nu$ corresponding to gaps in the bilayer spectrum
define the third, two-subband, Landau level fan as shown by solid
lines in Fig.~\ref{fan}. These lines are parallel to the ones of the
second subband fan so that with varying $V_g$ the bilayer electron
density changes essentially in the front part of the well. The
disruptions of the two subband fan lines at $\nu >2$ imply the
disappearance of common gaps for two subbands as will be discussed
below. As seen from Fig.~\ref{fan}, the two subband and first subband
fan lines for each $\nu=\nu_1$ intercept one another near the dotted
line at $V_g> V_{th}^2$. For description of all of these fan lines we
will use filling factor $\nu$ since $\nu_1$ at $V_g< V_{th}^2$ is
also determined by the electron density $N_s$ in the quantum well. A
part of the fan chart including the data points for the thermodynamic
density of states minima in the second subband is displayed in
Fig.~\ref{fan1}. At low magnetic fields the measurements are taken at
the lowest temperatures to obtain more pronounced capacitance minima
while at high $B$ the temperatures of about 1~K are necessary to
suppress the bilayer conductivity minima. These limits are indicated
by different symbols in the figure. The small deviations of the data
from the dashed lines give evidence that the intersubband electron
transfer is small and the unbalanced double layer system is soft.

\begin{figure}
\centerline{
\epsfxsize=18cm
\epsffile{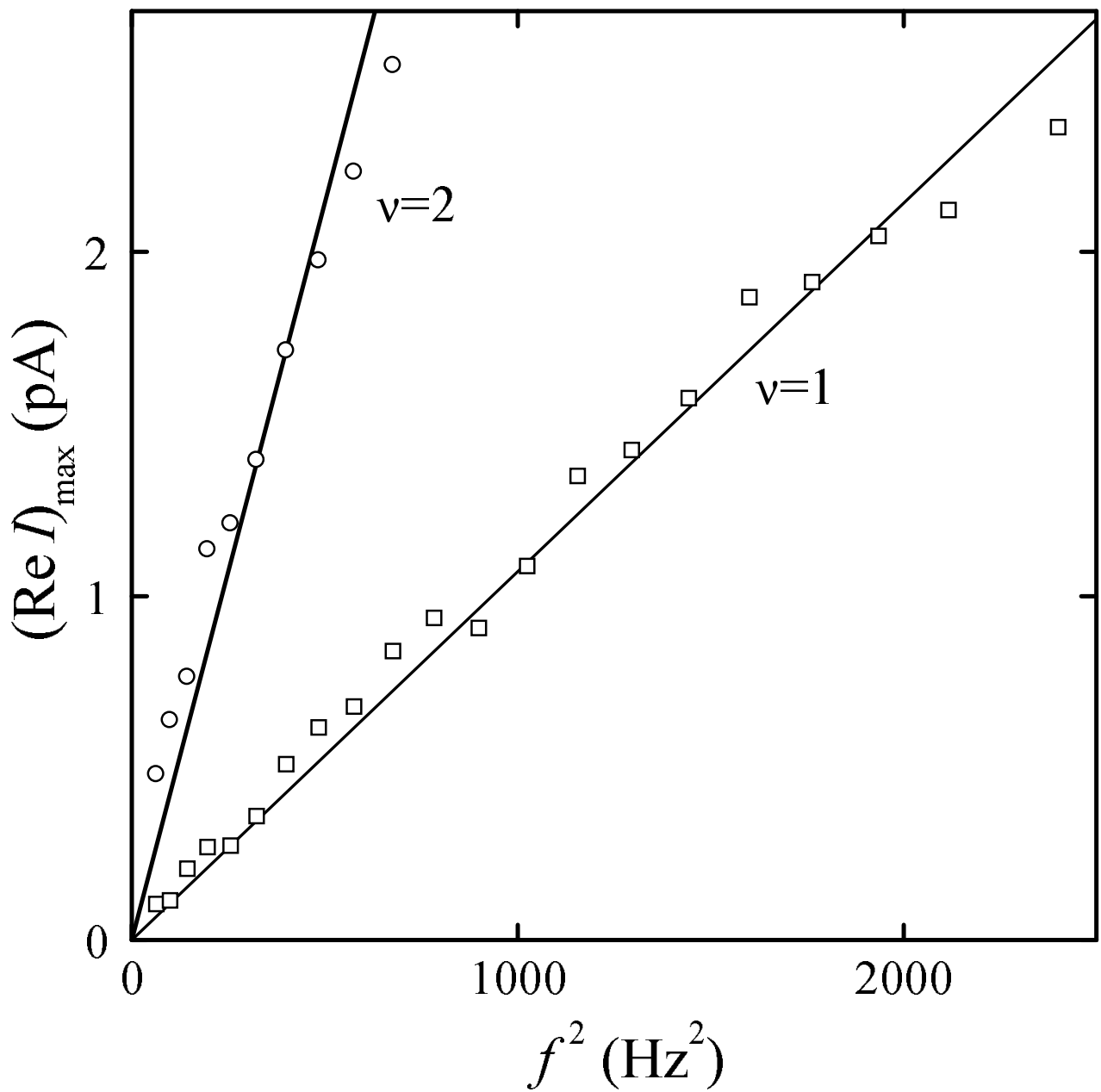}}
\caption{Frequency dependence of the active current component peaks
at $B=8$~T $T=0.62$~K.\label{peak}}
\end{figure}

Activation energy for gaps at integer $\nu$ is found from the
temperature dependence of peaks in the active current component. We
have checked that at sufficiently low frequencies the peak amplitude
is a parabolic function of $f$, see Fig.~\ref{peak}. As mentioned
above, the slopes of the straight lines in the figure should be
inversely proportional to $\sigma_{xx}$, which justifies the
procedure of activation energy determination from Arrhenius plot of
the peak amplitude.

\begin{figure}
\centerline{
\epsfxsize=18cm
\epsffile{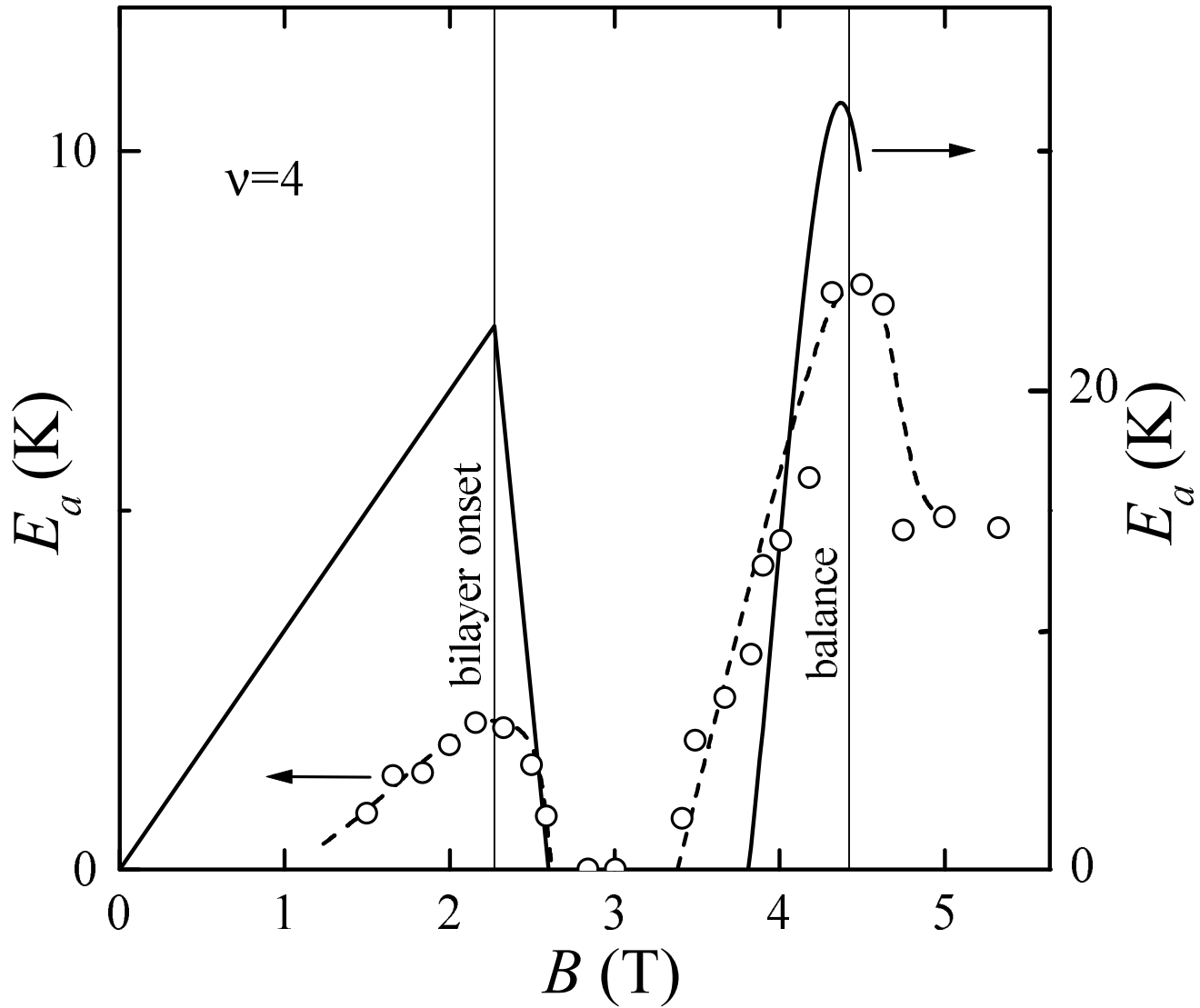}}
\caption{Experimental (dots) and calculated (solid line) changes of
the activation energy with magnetic field for filling factor $\nu=4$.
The dashed line is a guide to the eye.\label{ea4}}
\end{figure}

A typical behaviour of the activation energy $E_a$ with magnetic
field at fixed filling factor $\nu >2$ is depicted in Fig.~\ref{ea4}
for the case of $\nu=4$. The value of $E_a$ is a maximum both at the
bilayer onset $V_{th}^2$ and at balance. In between these it zeroes
in the interval 2.6 to 3.4~T except for the close vicinity of 3~T
where $E_a$ is unmeasurably small but likely finite as can be
reconciled with the non-zero active current component arising at a
fan crossing point $\nu=4,\nu_2=1$ ({\em cf.}
Figs.~\ref{fan},\ref{ea4}). The disruptions of the two subband fan
lines are the case between Landau level fan crossings
(Fig.~\ref{fan}) and correspond to the intervals of $B$ (or $V_g$) in
which the activation energy vanishes (Fig.~\ref{ea4}).

\begin{figure}
\centerline{
\epsfxsize=18cm
\epsffile{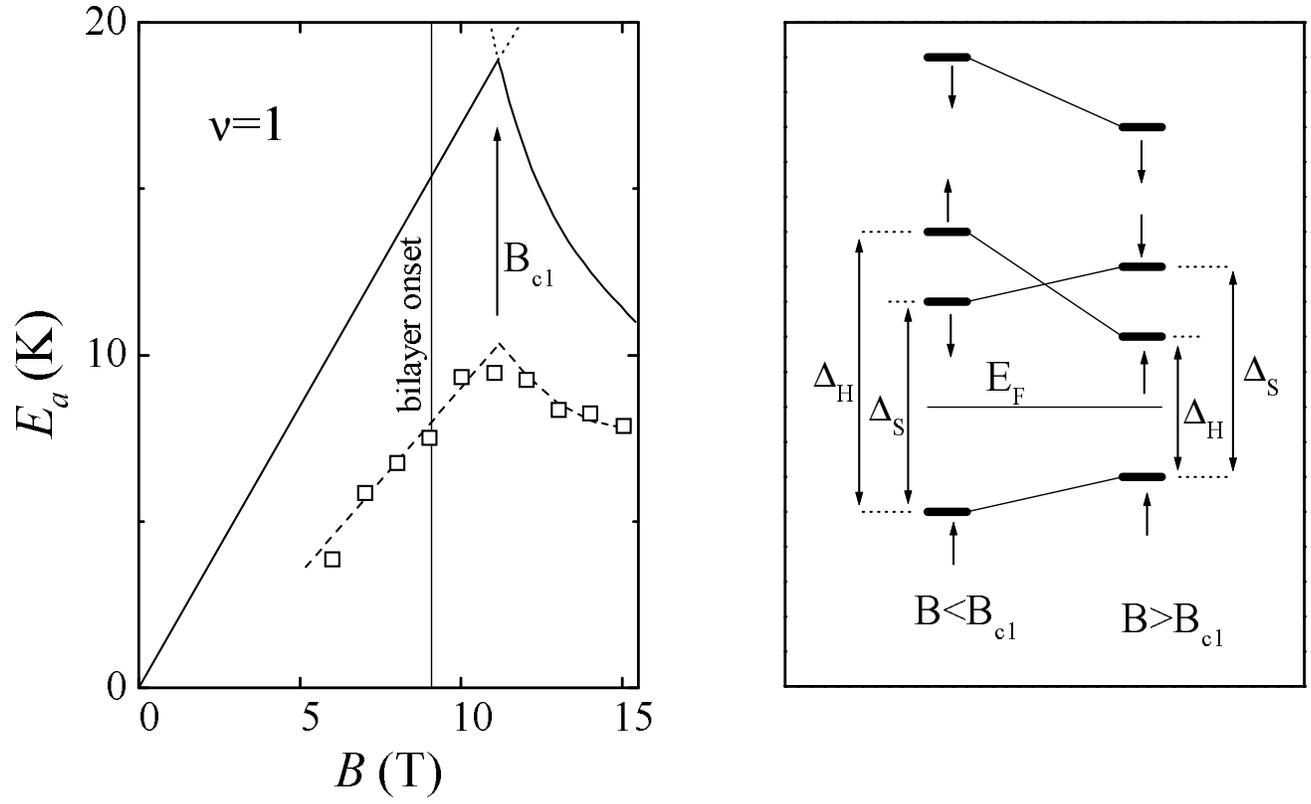}}
\caption{Activation energy as a function of magnetic field at
$\nu=1$: experiment (squares) and calculation (solid line). The
dashed line is a guide to the eye. Also shown is a sketch of the
bilayer spectrum below and above the maximum field
$B_{c1}$.\label{ea1}}
\end{figure}

In contrast, at $\nu=1,2$ the activation energy in the bilayer system
never tends to zero (Figs.~\ref{ea1},\ref{ea2}). For both filling
factors $E_a$ is a maximum near $V_g=V_{th}^2$ and then it
monotonically decreases with magnetic field up to the balance point.
We note that these $E_a$ maxima are attained at the interception
points of the two subband and first subband fan lines as described
above (Fig.~\ref{fan}).

\begin{figure}
\centerline{
\epsfxsize=18cm
\epsffile{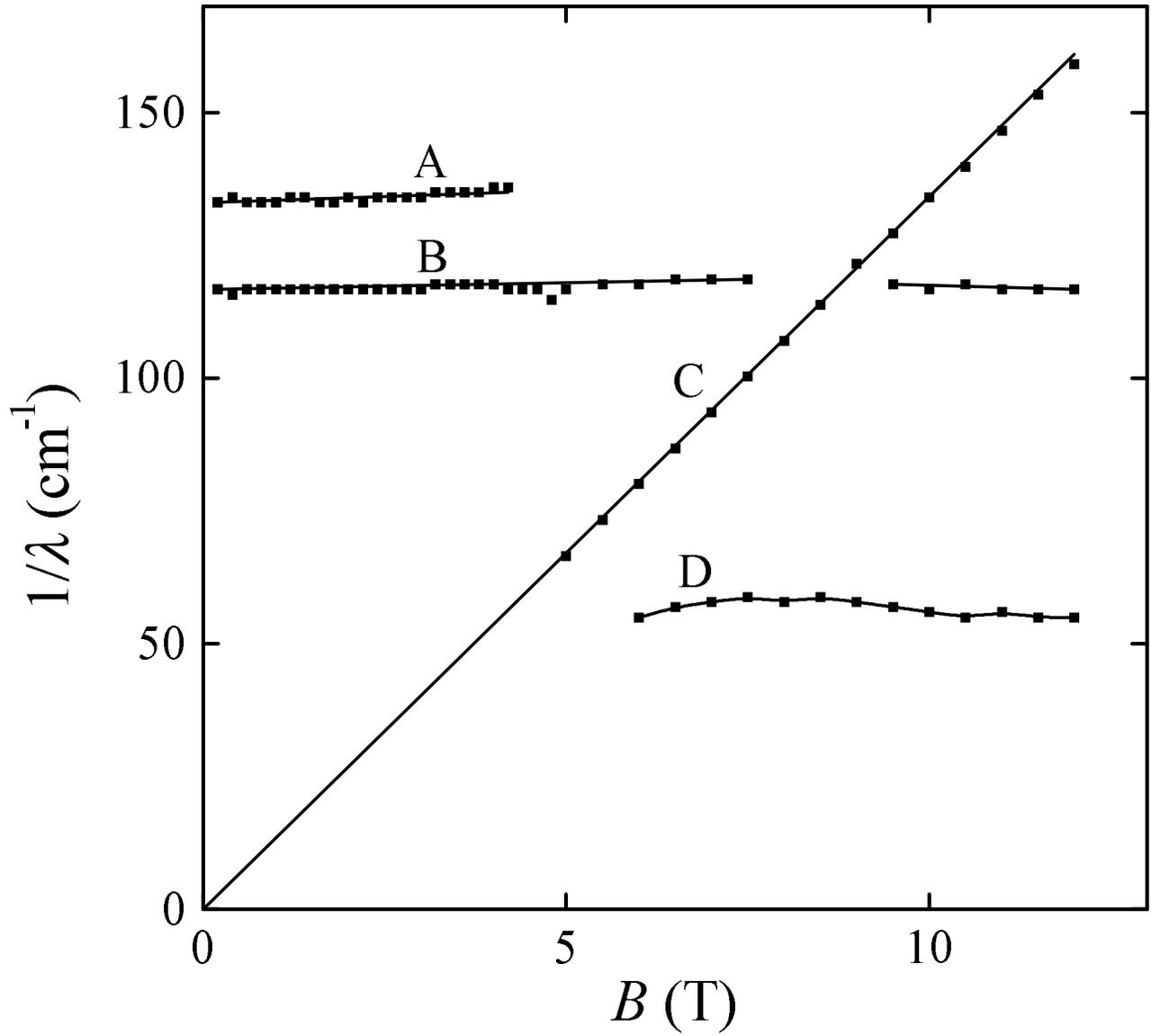}}
\caption{FIR absorption spectra in different magnetic fields at
$V_g=-0.15$~V.\label{fir}}
\end{figure}

Fig.~\ref{fir} shows FIR absorption spectra in magnetic field at
$V_g=-0.15$~V on our sample. Line C is easy to identify as cyclotron
resonance owing to the proportional $B$ dependence of its energy with
the slope as determined by an effective mass $0.07m_0$ ($m_0$ is the
free electron mass). The cyclotron resonance is observed with or
without grating coupler on the top of the sample. According to
Ref.~\cite{hart}, the rest three lines reflect the single-electron
spectrum in the quantum well, taking account of depolarization shift.
This implies that the screening-dependent subband energies in the
single-electron spectrum should be different for zero wave-vector
excitations. All of the three lines are seen only in the presence of
a grating coupler, which is in agreement with their intersubband
origin. Two lines A and B are attributed to electron transitions
between the second and the fourth subband and the first and the third
subband, respectively \cite{hart}. The most interesting line D
emerges at a magnetic field of 6~T corresponding to $\nu\approx 2$
and its energy is nearly independent of $B$ (Fig.~\ref{fir}). This is
the D line which we attribute to a hybrid gap in the double layer
spectrum.

\section{Discussion}
\label{disc}

\begin{figure}
\epsfysize=20cm
\epsffile{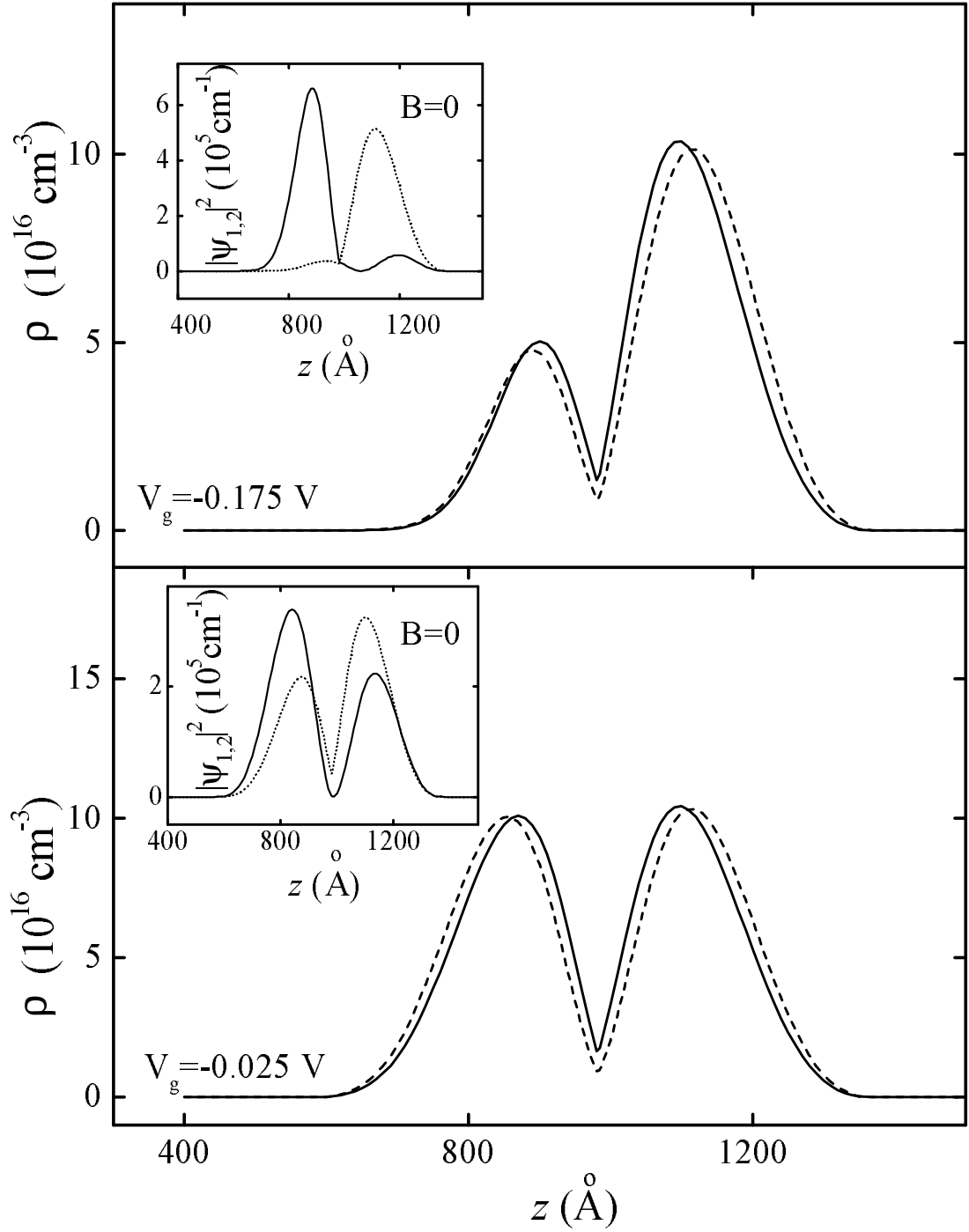}
\caption{Calculated electron density distributions in the quantum
well for two gate voltages at zero magnetic field (dashed lines) and
at filling factor $\nu=1$ (solid lines). The $z$ coordinate is
counted from the crystal surface. The electron density profiles for
both subbands at $B=0$ are shown in the insets.\label{prof}}
\end{figure}

The band structure of our sample in the absence of magnetic field is
known from far-infrared spectroscopy and magnetotransport
investigations on samples fabricated from the same wafer
\cite{hart1,ens}. It agrees with the result of self-consistent
Hartree calculation of energy levels in a coupled double quantum
well. According to the calculation, the symmetric-antisymmetric
splitting in the balanced bilayer system, which is the case at zero
gate voltage, is equal to $\Delta_{SAS}=1.3$~meV. With decreasing
$V_g$ the difference in the first and second subband energies should
enhance achieving the value of 6.7~meV at the bilayer onset. The
calculated electron density profiles, $|\psi_{1,2}|^2$, for both
subbands at two gate voltages are displayed in the insets to
Fig.~\ref{prof}. Because of tunneling the wave functions are not
localized in either part of the quantum well. The corresponding
electron density distributions in the quantum well $\rho=
N_s^1|\psi_1|^2 +N_s^2|\psi_2|^2$ (where $N_s^{1,2}$ are the
subbands' electron densities) are shown in Fig.~\ref{prof} by dashed
lines.

For an unbalanced double layer, switching a quantizing magnetic field
results in intersubband electron transfer, $n_s$, to minimize the
system energy that is accompanied by a relative shift of Landau level
ladders for two subbands. This shift can be estimated as

\begin{equation}
\Delta\approx 4\pi e^2n_sd/\varepsilon .\label{eq2}\end{equation}
The system softness $\alpha$ is characterised by the ratio of the
shift $\Delta/2$ of an individual ladder caused by transfer of all
electrons in the Landau level to the cyclotron energy, which is
consistent with the relation (\ref{eq1}). Given two Landau level
ladders, the origin of the third two-subband Landau level fan is
explained in the following way. At fixed integer $\nu$ the Fermi
level $E_F$ can be expected either to pin to both quantum levels for
two subbands or to fall within a common gap of the spectrum. The
latter should occur around the fan crossing points, at which both
$\nu_1$ and $\nu_2$ are integer, whereas in the pinning regions in
between the common gap should close. It is easy to obtain that the
ratio of the region dimensions where the common gap is absent and,
respectively, present is proportional to $\alpha$. Hence, in a
conventional two-subband system with vanishing $\alpha$ the pinning
regions should reduce to points and, oppositely, in a soft
two-subband system with $\alpha\agt 1$ these should extend up to fan
crossing points. The latter statement is in agreement with the data
at $\nu >2$ for our double layer in which $\alpha$ reaches 2.5, see
Figs.~\ref{fan},\ref{prof}. However, this crude approach which deals
with zero-magnetic-field subbands in an unbalanced double layer is
not good at filling factor $\nu=1,2$, at least (Fig.~\ref{fan}).

To obtain more rigorous solution of the problem in magnetic field we
start from the zero $B$ solution and determine the first order
corrections, $\Delta_{ij}$, to the subband energies $E_1$ and $E_2$,
caused by intersubband charge transfer in magnetic field

\begin{equation}
\Delta_{ij}=\frac{4\pi e^2}{\varepsilon}\int_{-\infty}^{\infty}
\Theta_{ik}^*(x,y)\Theta_{jl}(x,y){\rm d}x{\rm d}y
\psi_i^*(z)\psi_j(z){\rm d}z \int_{-\infty}^z {\rm d}z'
\int_{-\infty}^{z'} \Delta\rho(z''){\rm d}z'' ; i,j=1,2.\label{eq3}
\end{equation}
Here $\Theta_{ik}$ is the wave function of the level with Landau
quantum number $k$ in the $i$-th subband, which is closest to the
Fermi level, and $\Delta\rho$ is the variation of the electron
density distribution when switching a magnetic field. In our simple
model we will disregard exchange and correlation effects.

\begin{figure}
\centerline{
\epsfxsize=18cm
\epsffile{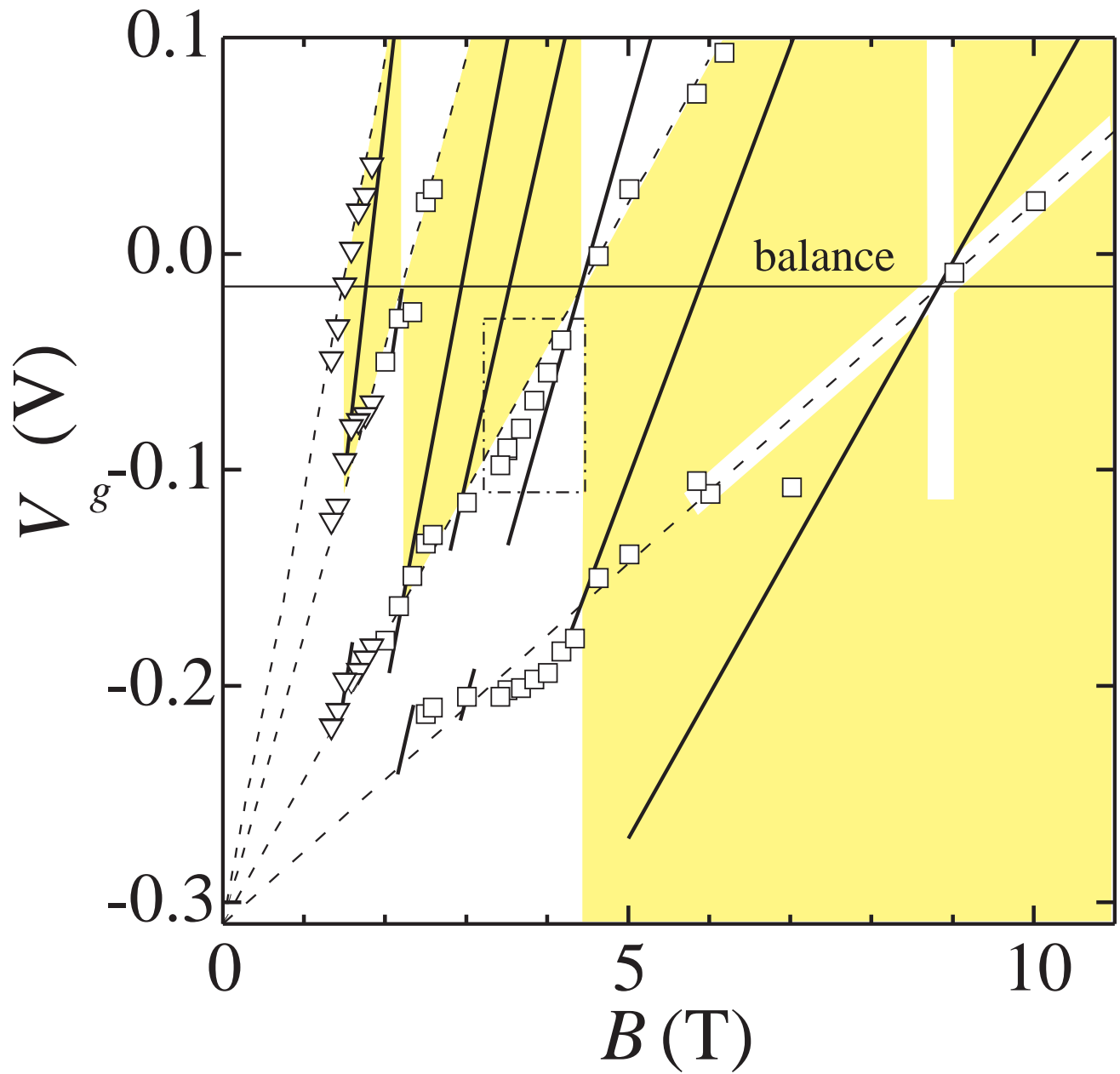}}
\caption{Part of the fan chart in Fig.~\protect\ref{fan}, including
the data points for density of states minima in the second subband.
The hybridization of electron subbands is expected in shaded
regions.\label{fan1}}
\end{figure}

In the region of common gaps for two subbands (unshaded area in
Fig.~\ref{fan1}) the quantum level numbers $k,l$ for $i\neq j$ are
different and so the off-diagonal $\Delta_{ij}$ are equal to zero,
leaving the wave functions $\psi_{1,2}$ unchanged. This region is
confined by the second subband fan lines and the expected vertical
fan lines for the first subband as shown in the figure. The value
$\Delta\rho$ here is given by

\begin{equation}
\Delta\rho=\left(|\psi_2|^2 -|\psi_1|^2\right)n_s ,\label{eq4}
\end{equation}
where $n_s$ is electron transfer from the second to the first
subband. One can see from Eqs.~(\ref{eq3},\ref{eq4}) that in the
general case the sum $\Delta_{11}+\Delta_{22}\neq 0$. At integer
$\nu$ the intersubband electron transfer gives rise to one of the
following: either both of the quantum levels in question reach the
Fermi level so that the common gap collapses or the top quantum level
becomes empty keeping a gap. Obviously, the latter occurs near the
fan crossing points. In the simplest case of equal gaps for two
subbands the common gap reduces by $|\Delta_{22}-\Delta_{11}|$.

The result of the common gap calculation at $\nu=4$ using the above
procedure is shown in Fig.~\ref{ea4}. To allow for the discrepancy
between the calculated and measured values of sample capacitance, the
$B$ values for the theoretical dependence $E_a(B)$ are multiplied by
a factor of 1.1 to fit the known electron density. For the sake of
simplicity the spin splitting is ignored and so there are only two
maxima on the expected dependence of the activation energy on
magnetic field. Although the qualitative agreement with the
experiment is good, the expected values of $E_a$ are very different
from the experimental data. This is likely to be due to a potential
disorder in the double layer that leads to the finite width of
quantum levels.

We note that if the Fermi level lies in a common gap, the
intersubband charge transfer is given by deviations of the
experimental data for the second subband fan from the straight lines
(see, e.g., the region marked by rectangle in Fig.~\ref{fan1}). That
these deviations are small excludes the possibility of all electrons
collecting in one part of the quantum well (so-called broken-symmetry
states \cite{jung}).

Over shaded areas in Fig.~\ref{fan1}, the Landau level numbers $k$
and $l$ for $i\neq j$ in Eq.~(\ref{eq3}) are coincident so that the
off-diagonal perturbation terms $\Delta_{ij}\neq 0$ and the solution
should be searched in the form of linear combination of $\psi_{1,2}$.
To simplify the problem, we will not mix states with antiparallel
spins, ignoring spin flip processes in tunneling. This is reasonable
once exchange and correlation effects are neglected. Then, in the
Hartree approximation one has to solve the determinant equation (for
certainty we consider the case of $k=l=0$)

\begin{equation}
{\rm det}\left(\begin{array}{cc}
\Delta_{11}+E_1-E_{1,2}^H&\Delta_{12}\\
\Delta_{21}&\Delta_{22}+E_2-E_{1,2}^H
\end{array}\right)=0 \label{eq5}\end{equation}
to find the hybrid subband energies $E_{1,2}^H$. These states are
described by the hybrid wave functions

\begin{equation}
\psi_{1,2}^H=\left(\Delta_{12}^2 +(\Delta_{11}+E_1-E_{1,2}^H)^2
\right)^{-1/2}\Big(\Delta_{12}\psi_1 -(\Delta_{11}+E_1-E_{1,2}^H)
\psi_2\Big) \label{eq6}\end{equation}
and thus we get for $\Delta\rho$ the following expression instead of
Eq.~(\ref{eq4})

\begin{equation}
\Delta\rho=(N_s^1+N_s^2)|\psi_1^H|^2 -N_s^1|\psi_1|^2
-N_s^2|\psi_2|^2 ,\label{eq7}\end{equation}
where $\psi_1^H$ corresponds to the lower energy state $E_1^H$.
Eqs.~(\ref{eq3},\ref{eq5}-\ref{eq7}) compose a perturbation theory
loop for self-consistent solution of the problem. As compared to a
common gap for two electron subbands, the hybrid gap $\Delta_H=
E_2^H-E_1^H <E_2-E_1$ never zeroes since it determines energy spacing
between the subbands that is expected at minimum to be equal to
$\Delta_{SAS}$. So, the magnetic-field-induced hybridization
generalizes the case of symmetric electron density distributions
corresponding to formation of $\Delta_{SAS}$.

It is interesting to compare how the zero-magnetic-field electron
density distribution in the quantum well changes for common and
hybrid gaps. From Eq.~(\ref{eq4}) it follows that in the case of a
common gap some charge is transferred between the front and back
parts of the well, i.e., $\Delta\rho(z)$ is close to an antisymmetric
function and so the corrections to the subband energies $E_1,E_2$ are
of opposite sign. In contrast, for a hybrid gap electrons are
displaced nearly symmetrically toward the well centre
(Fig.~\ref{prof}), resulting in the energy shifts of the same sign
$E_1^H >E_1$ and $E_2^H >E_2$.

Now in our model we introduce spin splitting of the hybrid levels.
Since the expected hybrid splitting $\Delta_H$ is comparable to the
many-body enhanced spin splitting $\Delta_S$, the energy spectrum
should be determined by their competition. For the simplest case of
$\nu=1$ the many-body enhanced spin gap as estimated for the
effective $g$ factor of 5.2 \cite{dens} is dominant over the range of
magnetic fields used except near the bilayer onset where the hybrid
splitting approaches half of the cyclotron energy. That stands to
reason, it is the smaller splitting that is the place at filling
factor $\nu=1$. Therefore, the theoretical dependence is a maximum at
a magnetic field $B_{c1}$ above the second subband threshold. This is
in reasonable agreement with the experiment (see
Figs.~\ref{ea1},\ref{fan}) as the actual value of effective $g$
factor may be smaller because of disorder.

\begin{figure}
\centerline{
\epsfxsize=18cm
\epsffile{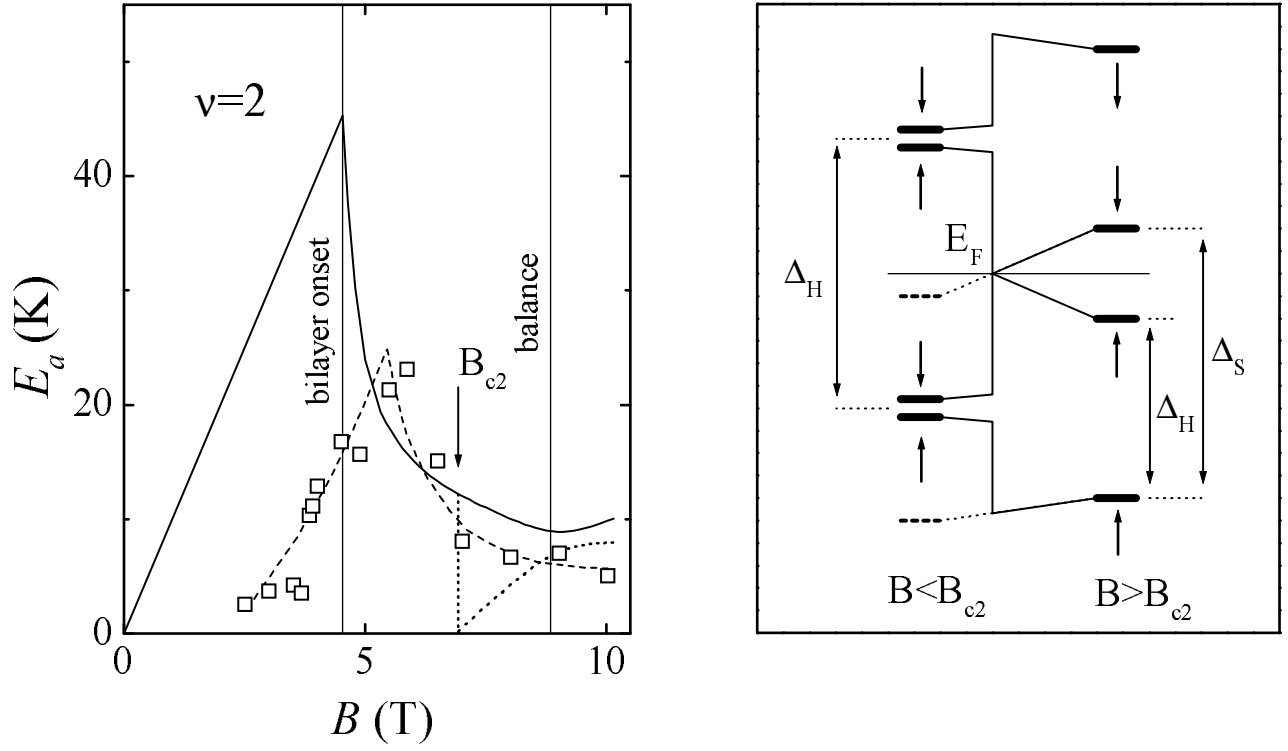}}
\caption{Comparison of the experimental (squares) and theoretical
(solid and dotted lines) dependences of the activation energy on
magnetic field for $\nu=2$. The dashed line is a guide to the eye.
The sketch displays the expected bilayer spectrum below and above the
critical field $B_{c2}$.\label{ea2}}
\end{figure}

For filling factor $\nu=2$ the situation is far more sophisticated
because, firstly, the actual gap is given by the splittings'
difference and, secondly, the $g$ factor can be enhanced or not,
dependent on filling of the spin sublevels. At least, near the
bilayer onset the hybrid gap exceeds the estimated many-body enhanced
spin gap. As long as in this case the spin sublevels with
antiparallel spins are filled, the spin splitting should correspond
to the $g$ factor in bulk GaAs $|g|=0.44$. This is true until the
hybrid and many-body enhanced spin splittings get equal at a higher
magnetic field $B_{c2}$ (Fig.~\ref{ea2}). If $B> B_{c2}$, it is
energetically more favourable for the system to have the spin
splitting enhanced so that the spin sublevels with parallel spins are
filled. Hence, the expected gap at $\nu=2$ should collapse around
$B=B_{c2}$ and then increase with magnetic field as shown in the
figure. Yet, this scenario is not realized in our double layer. This
implies that the actual many-body enhanced $g$ factor is smaller than
the one used in the calculation, i.e., the hybrid gap is dominant
over the entire range of fields up to the balance point. Therefore,
in the other scenario the gap at $\nu=2$ should merely be equal to
the difference between the hybrid and Zeeman splittings, which is in
qualitative agreement with the data (Fig.~\ref{ea2}). Thus, in our
experiment both gaps at $\nu=1,2$ are of hybrid origin as caused by
interplay of the hybrid and spin splittings. Apparently, for odd $\nu
>1$ near the balance one can expect gaps of spin origin that are the
smallest at lower magnetic fields. We note that the above scenarios
completely account for results of activation energy measurements in a
double layer system at filling factor $\nu=2$ of Ref.~\cite{saw}.

From comparison of Figs.~\ref{fan1},\ref{fir} it follows that line A
(D) is observed in the magnetic field range where the wave function
hybridization is absent (present). Switching these lines happens near
the shaded area boundary at $B=4.4$~T. This as well as the weak
magnetic field dependence of the energy of line D confirm its hybrid
origin.

Although adequate, our approach does not explain satisfactorily the
presence of data points at $\nu_2=1$ in the hybridization region
(Fig.~\ref{fan1}). In principle, peculiarities are indeed expected in
the vicinity of one subband fan lines for odd $\nu_1$ and $\nu_2$
since the intersubband electron transfer cannot occur there if
$E_2-E_1 <\Delta_S$. This inequality determines the regions where the
hybridization is absent and the subbands should be original (see the
white strips in Fig.~\ref{fan1}). In fact, for our case the above
inequality is never fulfilled as follows from the hybrid origin of
the gap at $\nu=2$ ({\em cf.} Figs.~\ref{fan1},\ref{ea2}). This
contradiction is not resolved within the simple model above and
demands consideration of exchange and correlation effects.

\section{Conclusion}
\label{conc}

In summary, we have performed magnetocapacitance and FIR measurements
on a bilayer electron system in a parabolic quantum well with a
narrow tunnel barrier in its centre. At imbalance created by gate
depletion we observe individual subband gaps and gaps in the double
layer spectrum at integer $\nu$. The latter are analyzed to be either
common for two subbands or of hybrid origin as caused by spectrum
reconstruction in magnetic field. With the help of a model for
magnetic-field-induced hybridization of electron subbands, we explain
the observed behaviour of the double layer gaps at $\nu=1$ and
$\nu=2$, making allowance for the competition between the hybrid and
spin splittings. Our approach is also capable of describing recent
results obtained elsewhere.

\acknowledgements

We gratefully acknowledge J.P.~Kotthaus, A.V.~Chaplik and M.~Shayegan
for fruitful discussions of the results. This work was supported in
part by  Deutsche Forschungsgemeinschaft, AFOSR under Grant
No.~F49620-94-1-0158, the Russian Foundation for Basic Research under
Grants No.~97-02-16829 and No.~98-02-16632, and the Programme
"Nanostructures" from the Russian Ministry of Sciences under Grant
No.~97-1024. The Munich - Santa Barbara collaboration has been also
supported by a joint NSF-European Grant and the Max-Planck research
award.

\end{document}